\pdfoutput=1
\documentclass[conference, a4paper]{IEEEtran}
\usepackage{times}
\usepackage{graphicx,color}
\usepackage{cite}
\usepackage{amsmath}
\newcommand{\bbar}[1]{\overline{\overline{#1}}{}}

\begin{document}

% Explicit linebreaks \\ can be added in the title. Do not put math or
% special symbols in the title.
\title{Electromagnetic wave propagation in\\ metamaterials: a visual guide to Fresnel-Kummer surfaces and their singular points}

\author{\IEEEauthorblockN{%
     Alberto Favaro\IEEEauthorrefmark{1}}
  \medskip
   \IEEEauthorblockA{\IEEEauthorrefmark{1}Department of Physics, Imperial College London, UK\\
    e-mail: a.favaro@imperial.ac.uk}}

\maketitle

% Do not use any symbols, special characters or math in the abstract.
\begin{abstract}
  The propagation of light through bianisotropic materials is studied in the geometrical optics approximation. For that purpose, we use the quartic general dispersion equation
  specified by the Tamm-Rubilar tensor, which is cubic in the
  electromagnetic response tensor of the medium. A collection of
  different and remarkable Fresnel (wave) surfaces is gathered,
  and unified via the projective geometry of Kummer surfaces.
\end{abstract}

\vspace{6pt}
\begin{figure}[h]
    \centering
    \includegraphics[width=0.75\columnwidth]{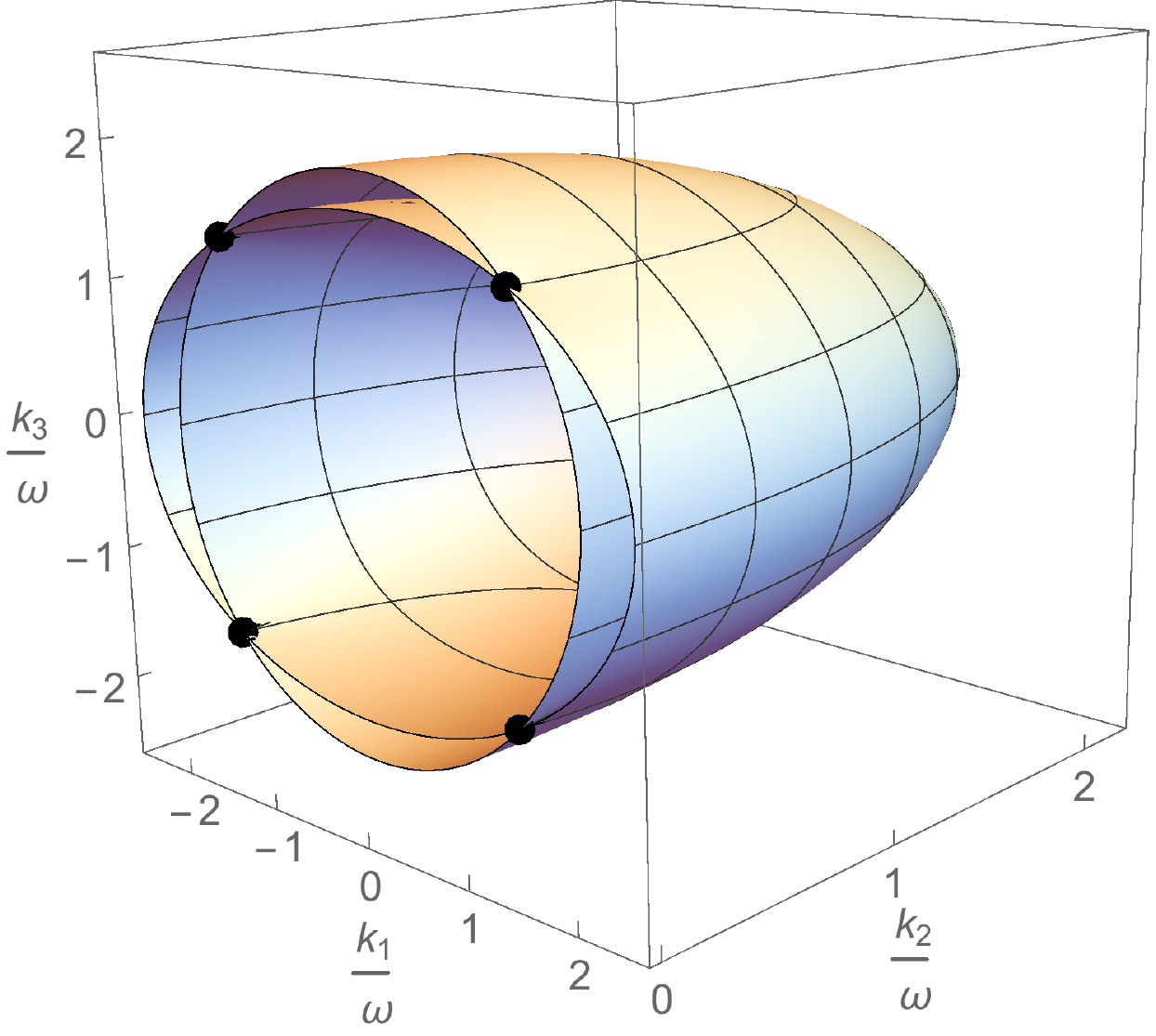}
    \caption{Cross section of the Fresnel surface for an ordinary biaxial
medium. (In this plot, the vacuum speed of light is set to one.)}
    \label{fig:biaxial}
\end{figure}
\section{Introduction}
\begin{figure}
    \centering
    \includegraphics[width=0.75\columnwidth]{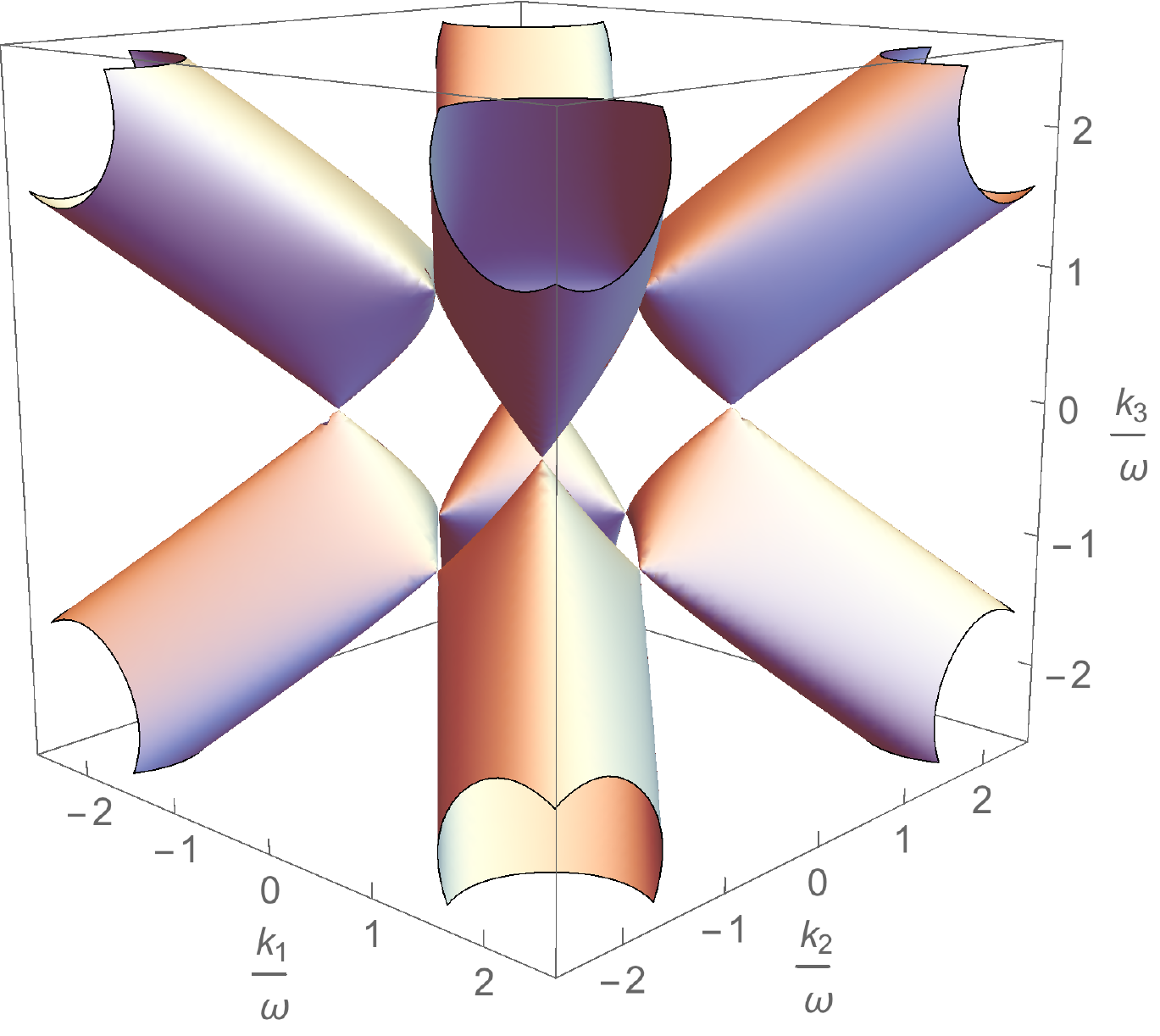}
    \caption{Fresnel surface with 12 singularities. The special points lie on the edges of a cube, at the midpoints. (The speed of light is set to one.)}
    \label{fig:inf4}
\end{figure}

\begin{figure}
    \centering
    \includegraphics[width=0.75\columnwidth]{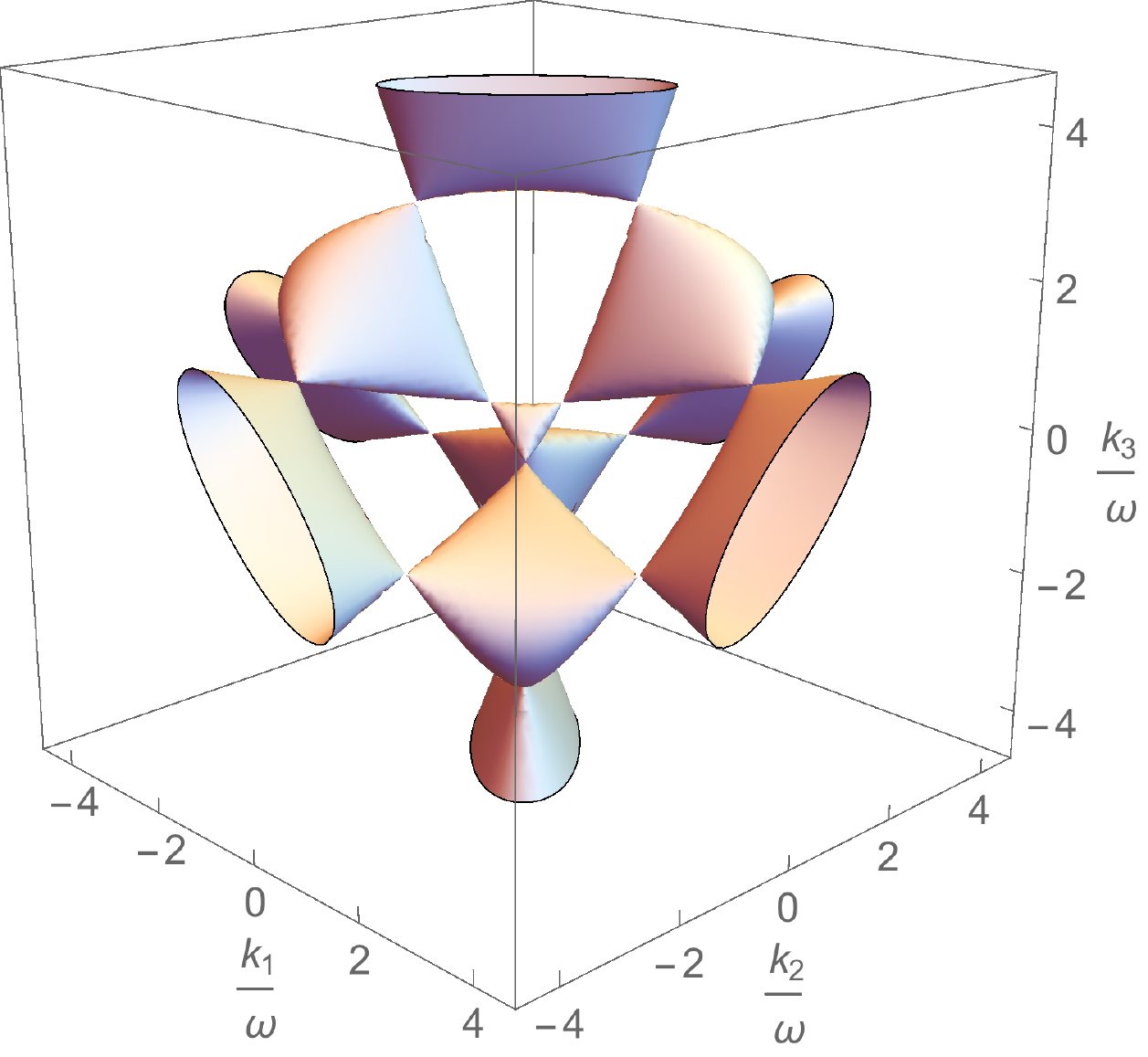}
    \caption{Fresnel surface with 16 singular points. This represents the general case of a Kummer surface. (The speed of light is set to one.)}
    \label{fig:noinf}
\end{figure}

The Fresnel wave surface of an electromagnetic medium with distinct anisotropic permittivities
has four isolated singular points (Figure \ref{fig:biaxial}). As soon as anisotropic
permeabilities and magnetoelectric terms are also considered, the
number of singularities can exceed four (Figure \ref{fig:inf4}). We propose a local and linear material
whose Fresnel surface displays sixteen singular points (Figure \ref{fig:noinf}). Because the general dispersion equation is quartic, this is the maximum allowed number of singularities \cite{Hudson}. To realize such extreme light-propagation geometries, one can take advantage of metamaterials. We suggest that mixtures of metal bars, split-ring resonators and magnetized particles yield the correct permittivities, permeabilities and magnetoelectric terms \cite{Favaro:2015jxa}. 

\section{Unusual and beautiful Fresnel surfaces}
\begin{figure}
    \centering
    \includegraphics[width=0.72\columnwidth]{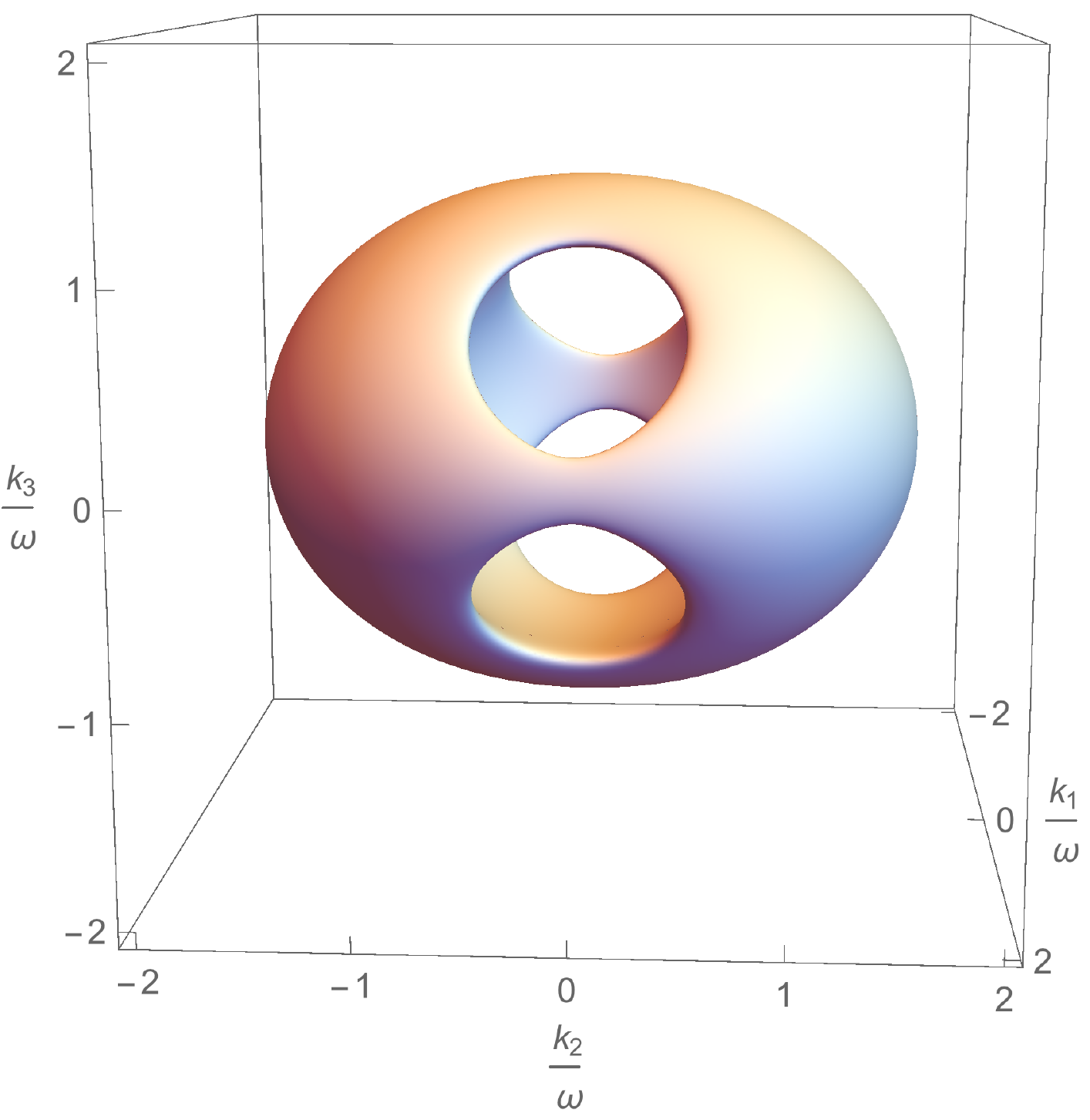}
    \caption{Fresnel surface of a medium with nonzero skewon part. Materials of this type are considered in \cite{Itin:2014gba, Obukhov:2004zz}. (The speed of light is set to one.)}
    \label{fig:skewonic}
\end{figure}

In this talk, uncommon and beautiful Fresnel surfaces are visualized and explored, both by resorting to the literature \cite{Dahl,Favaro:2011a,
  IsmoAri:2016,Obukhov:2004zz,Itin:2014gba}, and by generating
new images such as Figure \ref{fig:skewonic}. We also describe the physical specifications of the (meta)materials that correspond to the surfaces.

\section{Steiner's Roman Surface}
\begin{figure}
  \centering
    \includegraphics[width=0.77\columnwidth]{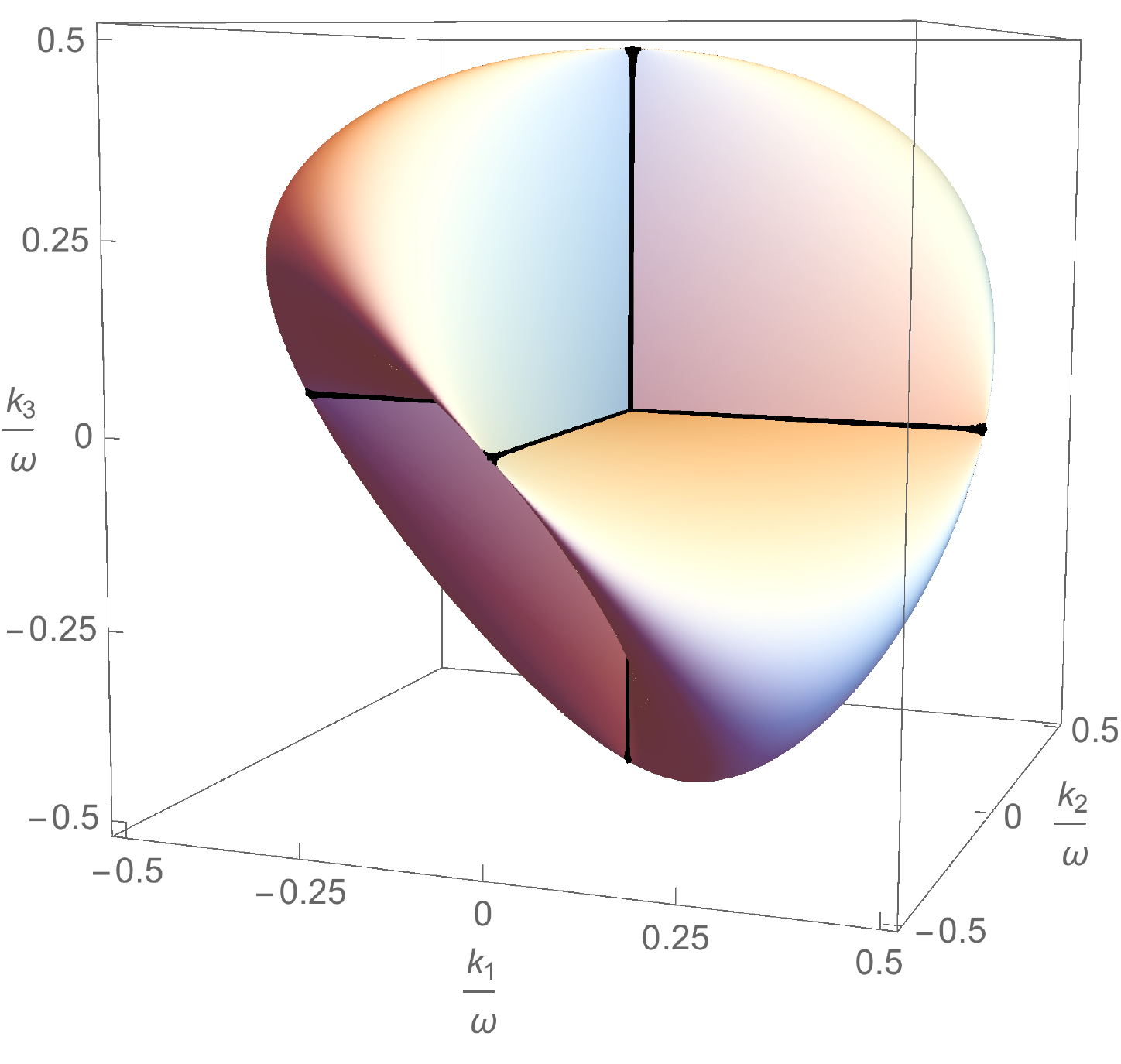}
    \caption{\label{fig:steiner} The Roman surface generated by \eqref{eq:roman} with $u=1$ and $c=1$.}
\end{figure}

As an example of unusual wave geometry, a linear material is proposed whose Fresnel surface coincides with the Roman surface. Figure \ref{fig:steiner} portrays this surface, whose name refers to the Italian city where it was invented by Swiss mathematician Jakob Steiner during a visit in 1844 \cite{Gray,Berger}. The Roman surface has delightful properties, for instance, it can be constructed from a M\"obius band and a disk by sewing the only edge of the former to that of the latter \cite{Francis}. For light propagation, this geometry is realized by the medium
%\end{equation}
\begin{equation}
\begin{pmatrix}
\mathbf{D}\\
\mathbf{H}
\end{pmatrix}
=%%
\begingroup
\renewcommand\arraystretch{1.34}
\left(\begin{array}{c|c}
-\bbar{\varepsilon} & \bbar{\alpha}\\[-1.5pt]\hline
\bbar{\alpha}^{\hspace{1.2pt}T} & \bbar{\mu}^{-1}
\end{array}\right)
\endgroup
\cdot%%
\begin{pmatrix}
-\mathbf{E}\\
\mathbf{B}
\end{pmatrix}\label{eq:magnetoelectric}
\end{equation}
with permittivity, permeability and magnetoelectric matrices
\begin{align}
\bbar\varepsilon&=0,\label{eq:epsilon}\\
\bbar\mu&=(u/40)^{\frac{2}{3}}\mu_{0}\hspace{0.7pt}\mbox{diag}(25,16,1),\label{eq:mu}\\
\bbar\alpha&=(u/40)^{\frac{1}{3}}(\varepsilon_0/\mu_0)^{\frac{1}{2}}\mbox{diag}(1,2,-3).\label{eq:alpha}
\end{align}
Here, $u$ is a nonvanishing parameter, while $\varepsilon_{0}$ and $\mu_{0}$ have the usual meaning. 

Even though the permittivity matrix is zero, the constitutive law \eqref{eq:magnetoelectric}--\eqref{eq:alpha} is invertible, so that one can write $\mathbf{E}$ and $\mathbf{B}$ as functions of $\mathbf{D}$ and $\mathbf{H}$. In fact, the determinant associated to the medium does not vanish, 
\begin{equation}
\det\hspace{-1pt}%%
\begingroup
\renewcommand\arraystretch{1.34}
\left(\begin{array}{c|c}
-\bbar{\varepsilon} & \bbar{\alpha}\\[-1.5pt]\hline
\bbar{\alpha}^{\hspace{1.2pt}T} & \bbar{\mu}^{-1}
\end{array}\right)
\endgroup
=-(3u/20)^{2}(\varepsilon_0/\mu_0)^{3}.\label{eq:determinant}
\end{equation}
Similarly, because the permeability matrix is invertible \cite{Ismo}, the constitutive law \eqref{eq:magnetoelectric}--\eqref{eq:alpha} can be reformulated with $\mathbf{D}$ and $\mathbf{B}$ in terms of $\mathbf{E}$ and $\mathbf{H}$. This way of pairing up the electromagnetic fields is standard among engineers. 

One can show that the dispersion equation of the medium being investigated reads
\begin{equation}
 -(\varepsilon_0^2/\mu_0^{3})\Bigl[k_{2}^2k_{3}^2+k_{3}^{2}
    k_{1}^{2}+k_{1}^{2}k_{2}^{2}-u(\omega/c) k_{1}k_{2}k_{3}\Bigr]=0, \label{eq:roman}
\end{equation}
where $\omega$ is the angular frequency, $\mathbf{k}=(k_1,k_2,k_3)$ is the wave vector, and $c=(\varepsilon_{0}\mu_{0})^{-\frac{1}{2}}$. Notably, \eqref{eq:roman} determines both the Roman surface of 
Figure \ref{fig:steiner}, and how light propagates inside the medium. Therefore, the geometry of Steiner's invention is reproduced by the electromagnetic waves.

By inspection, or by algebraic calculation, one verifies that the Roman surface in Figure \ref{fig:steiner} has three singular lines, namely the three coordinate axes. This makes the origin a peculiar kind of singularity called triple point. It is probable that, with regard to the motion of electromagnetic waves, the special lines and the triple point cause very interesting effects.

The decomposition of Hehl and Obukhov \cite{Birkbook} breaks up the material-law in three relativistic covariant parts. Most media consist only of a principal part. It is however possible that a material displays a nonvanishing skewon or axion part. For instance, measurements on Chromium Sesquioxide (Cr$_2$O$_3$) show that this crystal has a nonzero axion part, as well as a finite principal part \cite{Hehl:pra}. Even though the medium \eqref{eq:magnetoelectric}--\eqref{eq:alpha} yields the unusual Roman wave surface, its Hehl-Obukhov decomposition is rather ordinary. With the help of \eqref{eq:epsilon} and \eqref{eq:mu}, one verifies that the block matrix in \eqref{eq:magnetoelectric} is symmetric, so that the skewon part must be zero. Then, the magnetoelectric matrix \eqref{eq:alpha} is traceless, whereby the axion part vanishes. In summary, the constitutive law \eqref{eq:magnetoelectric}--\eqref{eq:alpha} exhibits no more than a principal part. 

As discussed in \cite{Rowe}, the Roman surface is a special example of Kummer surface. This can be regarded as a consequence of the identity between Fresnel and Kummer surfaces, which holds true for all media having zero skewon part (see below).    

\section{Tamm-Rubilar tensor and Kummer surfaces}
Fresnel surfaces are determined by a general Tamm-Rubilar tensor, cubic in the medium parameters. We plan to review the corresponding literature \cite{PhD,Ismo,Favaro:2015b,Birkbook,Itin:2009aa}, and to demonstrate
how this tensor brings order into the zoo of different wave surfaces for electromagnetic materials.

Another unifying principle is the identity of Fresnel and Kummer surfaces \cite{Baekler:2014kha,Bateman}, which can be described as follows: Fresnel surfaces of local and linear materials with zero skewon part are, with regard to projective geometry, Kummer surfaces. Moreover, every Kummer surface can be interpreted as the Fresnel surface of a medium with vanishing skewon part. 

This equivalence allows one to make an important observation about the singular points. It is known from classic projective geometry that Kummer surfaces have exactly sixteen isolated singularities over the complex numbers \cite{Jessop, Lord}. The same is then valid for the Fresnel surfaces of materials with zero skewon part. One should take note of the requirement to work over the complex numbers. The surface in Figure \nolinebreak\ref{fig:noinf} has sixteen real singular points, which can be visualized through rotations. By contrast, the surface in Figure \ref{fig:biaxial} has twelve complex singular points, which are impossible to depict. This explains why only four real singularities can be seen in the plot. A subtlety concerning singular points that are not finite is discussed in \cite{Favaro:2015jxa}, but neglected here. Future research will investigate the physical meaning of complex singularities on Fresnel surfaces.

\section*{Acknowledgments}
The author would like to thank Friedrich W.\ Hehl for instructive discussions. Financial support from the Gordon and Betty Moore Foundation is gratefully acknowledged.\\
 
% Shrink the final page to force the columns to be balanced. This
% command must be in the first column of the last page. Start by
% setting the value to 0mm and slowly increase it until the columns
% balance.
%\enlargethispage{-154mm}
\bibliographystyle{IEEEtran}
\bibliography{IEEEabrv,IEEEexample}

%\end{document}
\end{document}